\documentclass{article}
\usepackage{dcase2026_techrep,amsmath,amssymb,graphicx,url,times,booktabs,tabularx}
\usepackage{tikz}
\usepackage{stfloats}
\usetikzlibrary{arrows.meta,positioning,fit,calc}

\makeatletter
\renewcommand{\twoauthors}[4]{\gdef\@address{}
   \gdef\@name{\begin{minipage}[t]{0.5\textwidth}\begin{center}\begin{tabular}[t]{@{}c@{}}
        {\em #1} \\ \\
        #2\relax
   \end{tabular}\end{center}\end{minipage}\hfill\begin{minipage}[t]{0.5\textwidth}\begin{center}\begin{tabular}[t]{@{}c@{}}
        {\em #3} \\ \\
        #4\relax
\end{tabular}\end{center}\end{minipage}}}
\makeatother

\title{DOMAIN-INCREMENTAL AUDIO CLASSIFICATION USING DOMAIN-SPECIFIC  EXPERTS AND PROTOTYPE CLASSIFIER}

\twoauthors
  {\begin{tabular}[t]{@{}c@{}}
   Jongyeon Park$^{1}$, Do-Hyeon Lim$^{1}$, Sang-won Park$^{2}$,\\
   Hong Kook Kim$^{1,2,*}$\thanks{$^{*}$\,This work was supported in part by Hanwha Vision Co. Ltd. and by the ‘Science and Technology Opens the Future of the Region' program funded by MSIT and Gwangju Metropolitan City in 2026.}
   \end{tabular}}
  {$^1$Dept. of AI Convergence,
   $^2$Dept. of EECS,\\
   Gwangju Institute of Science and Technology\\
   Gwangju 61005, Korea\\
   \{\{jypark3737, do-hyeon, sangwonpark\}@gm.,\\
   hongkook@\}gist.ac.kr}
  {\begin{tabular}[t]{@{}c@{}}
   Kyungdeuk Ko$^{3}$, Hyeongcheol Geum$^{3}$,\\ Jeong Eun Lim$^{3}$
   \end{tabular}}
  {$^3$AI Lab., R\&D Center\\
  Hanwha Vision\\
   Seongnam-si, Gyeonggi-do 13488, Korea\\
   \{kd.kyung, hch.geum, je04.lim\}@hanwha.com}

\begin{document}

\ninept
\maketitle

\begin{sloppy}

\begin{abstract}
This technical report presents submission systems for Task 7 (domain-incremental audio classification) of the DCASE 2026 Challenge.
The main obstacle is that the system can never access past- and future-domain data at the same time. We approached domain-incremental learning (DIL) as a frozen-feature replay problem. At each incremental stage, one or two compact experts are trained and then kept fixed; at the final stage, the penultimate features from all frozen experts are concatenated and used to train a lightweight per-class prototype classifier solely on cached features. This design prevents catastrophic forgetting by preserving each frozen expert at inference.
To retain earlier-domain knowledge without raw audio, each expert is trained with DeepInversion-based generative replay. Separately, a cross-stage regression imputer---trained only on samples for which all expert slots are legitimately observable---fills the feature slots of experts that did not yet exist at an earlier stage. We submit four fully DIL-compliant systems: three based on diverse frozen five-expert backbones and their cross-stack ensemble, achieving 78.38\% micro / 78.92\% macro on the development set, outperforming every individual backbone on both metrics.
\end{abstract}


\section{Introduction}
\label{sec:intro}

DCASE 2026 Task 7~\cite{casciotti2026dcasetask7} presents a domain-incremental learning (DIL) problem in which three domains arrive in sequence: $D_1$ (audio not provided), $D_2$ and $D_3$ (audio provided). The model is evaluated after all three stages over a fixed set of 10 target classes (\textit{alarm}, \textit{baby\_cry}, \textit{bark}, \textit{engine}, \textit{fire}, \textit{footsteps}, \textit{knock}, \textit{telephone\_ringing}, \textit{piano}, \textit{speech}). For the model to perform ideally, it must not forget earlier domains while learning subsequent domains.
The training procedure is limited by DIL-compliance: at stage $k$, one may use only the raw audio of the current domain $D_k$, together with any model or feature already produced and stored at stages $\le k$.

In domain-incremental learning, a model must adapt to newly arriving domains while preserving its performance on previously learned domains. However, updating a shared model only with current-domain data can cause catastrophic forgetting\cite{forgetting}, where the model becomes biased toward the recent domain and degrades its performance on earlier domains.

Our submission incorporates three enhancements to address the challenges in DIL mentioned above. First, catastrophic forgetting could be mitigated by utilizing multiple domain-specific expert models\cite{domainexpert}. Each domain expert contributes complementary diversity to the overall model, enhancing robustness while preventing interference between the parameters of different domains.  Second, to effectively preserve past information without accessing the original domain data, we leverage DeepInversion\cite{yin2020deepinversion} to generate synthetic data approximating the previous domain and incorporate it into training. Finally, in order to combine multiple heterogeneous domain experts, we utilized a cosine prototype head based on prototype learning\cite{prototpye}. The cosine prototype head is trained on the cached penultimate features of each domain expert. Since domain-incremental learning prevents access to future-domain experts during earlier training stages, the resulting missing feature dimensions were estimated through cross-stage regression and used to construct the prototype representations.

Following this introduction,  Section ~\ref{sec:method} describes the proposed system architecture and methods. Later, Section ~\ref{sec:exp}  discusses the experimental results and Section~\ref{sec:concl} concludes the report.

\section{Proposed Method}
\label{sec:method}

\subsection{System composition}
\label{ssec:composition}

A system consists of a 3-seed bag\cite{3seed} of prototype classifiers and five domain-specific experts. The pipeline is shown in Fig.~\ref{fig:pipeline}. Raw audio is converted to a log-mel spectrogram and passed through five frozen experts.
Their penultimate vectors are concatenated ($5\times2048=10240$-d) and fed to a per-class prototype classifier whose 3-seed softmax outputs are averaged. Experts are the only components trained on audio; the prototype is trained purely on cached features.

\begin{figure}[t]
  \centering
  \resizebox{0.92\columnwidth}{!}{%
  \begin{tikzpicture}[
      node distance=3.2mm,
      box/.style={draw, rounded corners=1pt, align=center, font=\scriptsize,
                  minimum height=5.2mm, inner sep=2pt, fill=white},
      frz/.style={box, fill=black!4, minimum width=0.15\columnwidth},
      line/.style={thin},
      arr/.style={-{Latex[length=1.4mm]}, thin}]
    \node[box] (raw) {raw audio (32\,kHz, 4\,s)};
    \node[box, below=of raw] (mel) {log-mel (64 bins)};
    \node[frz, below=4.5mm of mel] (e3) {$E_3$};
    \node[frz, left=of e3]  (e2) {$E_2$};
    \node[frz, left=of e2]  (e1) {$E_1$};
    \node[frz, right=of e3] (e4) {$E_4$};
    \node[frz, right=of e4] (e5) {$E_5$};
    \node[box, below=4.5mm of e3, minimum width=0.6\columnwidth] (concat)
        {concat penultimate ($5\times2048=10240$-d)};
    \node[box, below=of concat, minimum width=0.6\columnwidth] (cb)
        {prototype classifier (per-class prototypes) $\times$3 seeds};
    \node[box, below=of cb] (out) {class probabilities (10), seed-averaged};

    \draw[arr] (raw)--(mel);

    \coordinate (tbus) at ($(e3.north)+(0,3mm)$);
    \draw[line] (mel.south) -- (mel.south |- tbus);
    \draw[line] (e1.north |- tbus) -- (e5.north |- tbus);
    \foreach \e in {e1,e2,e3,e4,e5}{\draw[arr] (\e.north |- tbus) -- (\e.north);}

    \coordinate (bbus) at ($(e3.south)+(0,-2mm)$);
    \foreach \e in {e1,e2,e3,e4,e5}{\draw[line] (\e.south) -- (\e.south |- bbus);}
    \draw[line] (e1.south |- bbus) -- (e5.south |- bbus);
    \draw[arr] (concat.north |- bbus) -- (concat.north);

    \draw[arr] (concat)--(cb);
    \draw[arr] (cb)--(out);
  \end{tikzpicture}}
  \caption{Stage-3 inference pipeline. Frozen experts are trained on audio; the prototype head is trained only on cached features.}
  \label{fig:pipeline}
\end{figure}
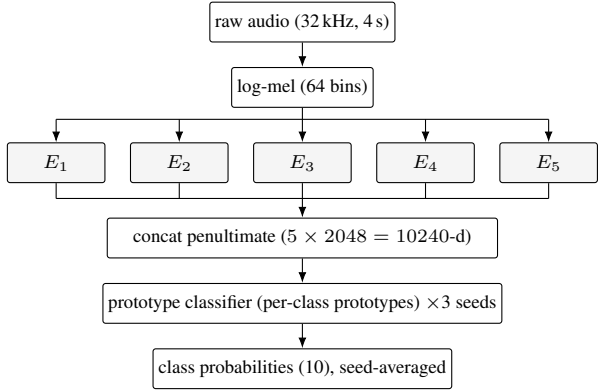

\begin{table}[t]
  \centering
  \caption{The three frozen five-expert backbones.}
  \label{tab:backbones}
  \resizebox{\columnwidth}{!}{%
  \begin{tabular}{@{}llllll@{}}
    \toprule
    Backbone & $E_1$ & $E_2$ & $E_3$ & $E_4$ & $E_5$ \\
    \midrule
    System 1& base-$D_1$ & base-$D_2$ & FDY-CNN14 $D_2$ & base-$D_3$ & scratch-$D_3$ \\
    System 2& base-$D_1$ & inv-$D_2$ & CRNN-light $D_2$ & inv-$D_3$ & scratch-$D_3$ \\
    System 3& base-$D_1$ & inv-$D_2$ & FDY-CNN14 $D_2$ & base-$D_3$ &  scratch-$D_3$ \\
    \bottomrule
  \end{tabular}}
\end{table}

\subsection{Five-expert backbones}
\label{ssec:backbones}

Before feature concatenation and prototype classification, the input log-mel spectrogram is independently processed by five frozen domain experts. The composition of these five-expert stacks differs across the three submitted systems, as summarized in Table~\ref{tab:backbones}. Across all systems, $E_1$, $E_2$, and $E_4$ are built upon the CNN14 architecture \cite{kong2020panns} and are trained on $D_1$, $D_2$, and $D_3$, respectively. The remaining $E_3$ and $E_5$ slots are populated with additional $D_2$ and $D_3$ experts using alternative architectures selected empirically.

\noindent\textbf{(1) Incremental domain experts.}  
To retain as much knowledge from $D_1$ as possible while extending the system's capability to the later domains, we first trained the experts base-$D_2$ and base-$D_3$ via sequential fine-tuning, initializing each expert from its predecessor in the training sequence. However, such a strategy is prone to catastrophic forgetting, which can lead to the loss of knowledge acquired from previous domains. To mitigate this problem, we additionally trained DeepInversion-based generative-replay variants (inv-$D_2$, inv-$D_3$); each submitted system uses either the base- or inv- variant(Table~\ref{tab:backbones}). Specifically, synthetic samples are generated from the previously trained models and replayed at subsequent training stage. This strategy enables experts to retain knowledge from previous domain while adapting to new domain.

\noindent\textbf{(2) Purely-trained domain experts.}
To enrich the feature representations for the prototype classifier, we augment the expert stack with two additional domain experts --- $E_3$ and $E_5$. These experts are trained from scratch on individual domains, corresponding to $D_2$ and $D_3$, respectively. For $E_3$, Systems 1 and 3 employ FDY-CNN14\cite{nam2022fdy}, while System 2 employs CRNN-light. In contrast, all systems use the same CNN14-based model for $E_5$. To reduce confusion among classes that exhibit similar acoustic characteristics, $E_5$ is trained with label smoothing(0.15)\cite{szegedy2016labelsmooth} and semi-hard negative mining\cite{schroff2015facenet}. These additional experts provide complementary feature representations that improve the discriminative capability of the prototype classifier. We added up to two extra experts(total five experts), since adding more experts yielded only marginal performance gains while incurring a substantially larger memory footprint.

\subsection{DeepInversion-driven generative replay}
\label{ssec:inversion}
Among various approaches such as regularization-based or architecture-based methods\cite{cl-review}, replay-based methods have consistently shown strong effectiveness in mitigating catastrophic forgetting.

Unlike conventional replay-based approaches that store and reuse raw audio samples from previously learned domains, we employ replay using synthetic data generated by DeepInversion\cite{yin2020deepinversion}. For each trained domain expert, we freeze the model and synthesize class-conditional log-mel features from randomly initialized inputs. During synthesis, only the input log-mel features are optimized using cross-entropy and BatchNorm-statistics matching losses.  These objectives encourage the synthetic features to reflect the domain knowledge from the frozen expert, enabling experts to train subsequent domain without storing raw audio from previous domains. 

Specifically, we define separate batch sizes, denoted by $B_S$ and $B_I$ ($B_I$ $<$ $B_S$), for the current domain dataset $S$ and the synthetic dataset $I$, respectively. The final training batch is then constructed by combining $B_S$ current-domain samples and $B_I$ synthetic samples, ensuring that both current and replayed data are exposed during training. 

\begin{figure}[t]
  \centering
  \begin{tikzpicture}[
    font=\footnotesize, >=Stealth, node distance=6mm and 9mm,
    box/.style={draw, align=center, minimum height=8mm, inner sep=3pt, text width=26mm},
    src/.style={box, rounded corners=2pt, fill=gray!12},
    feat/.style={box, fill=white},
    reg/.style={box, thick, fill=gray!20},
    cb/.style={draw, very thick, align=center, inner sep=4pt, text width=52mm, fill=gray!8},
    flow/.style={->, thick},
    learn/.style={->, thick, dashed},
  ]
    \node[src] (d2) {D2 audio $\to$ E1--E3\\(3 experts)};
    \node[src, right=of d2] (d3) {D3 audio $\to$ E1--E5\\(5 experts)};
    \node[reg, below=of d2] (reg) {Regressor $R$\\{\scriptsize$[\text{E1,E2,E3}]\!\to\![\text{E4,E5}]$}};
    \node[feat, below=of d3] (d3f) {Complete feats\\real E1--E5};
    \node[feat, below=of reg] (d2i) {Imputed feats\\real E1--E3, $\hat{\text{E4}},\hat{\text{E5}}$};
    \node[cb, anchor=north, xshift=17mm] (cb) at ([yshift=-7mm]d2i.south)
      {Prototype classifier training\\D2(imputed) $\cup$ D3(real)};
    \draw[flow] (d2) -- (reg);
    \draw[flow] (d3) -- (d3f);
    \draw[learn] (reg) -- node[right, font=\scriptsize, pos=0.5]{impute} (d2i);
    \draw[learn] (d3f.west) -- node[above, font=\scriptsize]{train $R$} (reg.east);
    \draw[flow] (d2i.south) -- (d2i.south |- cb.north);
    \draw[flow] (d3f.south) -- (d3f.south |- cb.north);
  \end{tikzpicture}
  \caption{Stage-compatible feature cache and regression imputation. The regressor $R$, trained only on the D3 row where all five expert slots are observable, predicts the missing E4,E5 slots for the D2 row.}
  \label{fig:impute-flow}
\end{figure}
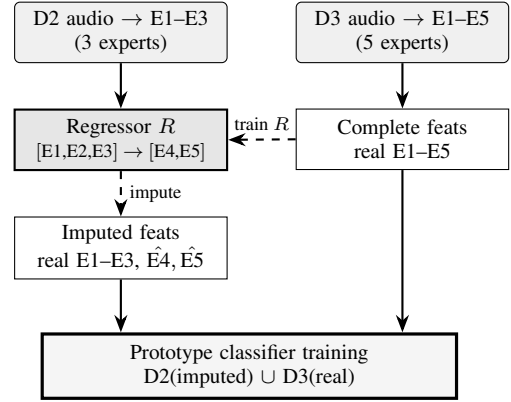
\subsection{Prototype classifier}
\label{ssec:prototype}

The prototype classifier head replaces a conventional softmax-linear head with a per-class prototype nearest-neighbor classifier in a per-expert-normalized space. For feature vectors $f\in\mathbb{R}^{10240}$ (concatenation of five 2048-d penultimate vectors):

\noindent\textbf{(1) Per-expert L2-normalization.}  
L2-normalization is applied to each of the five 2048-d chunks independently, $g_i = f_i/\lVert f_i\rVert_2$. The five experts are heterogeneous (CNN conv-pool features vs.\ CRNN BiLSTM features vs.\ Synthetic-replayed CNN features) and their penultimate vectors live on very different magnitude scales; per-expert normalization puts them on a common scale so that the concatenated cosine score weights each expert equally rather than letting the largest-norm expert dominate.

\noindent\textbf{(2) Concatenate and score.} Concatenate $g=[g_1,\dots,g_5]\in\mathbb{R}^{10240}$. The classifier holds one learnable prototype per class, $P\in\mathbb{R}^{10\times10240}$. The per-class score is the cosine similarity $s_c=\cos(g,P_c)$.

\noindent\textbf{(3) Temperature scaling and softmax.}
Since $s_c$ is a cosine similarity, it is bounded, $s_c\in\mathbb[-1,1]$. A softmax taken directly over this narrow range is nearly uniform. The cross-entropy 
is bounded away from zero and its gradients vanish~\cite{wang2017normface}. We therefore divide the scores by a learnable temperature, $p=\mathrm{softmax}(s/\tau)$, with $\tau$ initialized to $0.1$; during training it converges to $\approx0.005\text{--}0.008$ (an effective scale $1/\tau\approx125\text{--}200$), sharpening the bounded cosine scores into a usable posterior.

\subsection{Missing feature imputation}
\label{ssec:missing}

Prototype classifier requires the full five-expert feature vector of every clip, but a $D_2$ clip can only pass through experts that exist at stage~2, i.e.\ $E_1-E_3$; the $D_3$-stage experts $E_4,E_5$ do not exist at stage~2, and by stage~3 the $D_2$ raw audio is gone. Therefore, slots $E_4,E_5$ of every $D_2$ clip are unobservable, whereas $D_3$ clips have full feature vectors. $D_2$ rows hold real features in slots 1,2,3 and zeros in slots 4,5; $D_3$ rows are complete.

We train a 2-layer MLP regressor $R:[E_1,E_2,E_3](6144d) \xrightarrow{} [E_4,E_5](4096d)$ utilizing $D_3$ rows only. Then $R$ is applied to every $D_2$ row to fill its missing features: $E_4,E_5$. Both rows have identical width (10240), thus the prototype head input dimension is constant. This enables the model to learn how a clip's $D_1$/$D_2$-expert features relate to its $D_3$-expert features and extrapolates that relationship to $D_2$ clips.

\begin{table}[t]
  \centering
  \caption{Accuracy of the four submitted systems across domains D2 and D3. (Micro acc / Macro acc)}
  \label{tab:systems}
  \resizebox{\columnwidth}{!}{
  \begin{tabular}{lllc@{}}
    \toprule
     \#&  $D_2$&$D_3$&Dev\_test Avg \\\midrule
     Official checkpoint&  54.77 / 58.95&36.23 / 47.34&45.50 / 53.15\\
     System 1&   81.22 / 82.50&73.33 / 73.12&77.27 / 77.81 \\
     System 2&   79.97 / 81.88&75.19 / 73.97&77.58 / 77.92 \\
     System 3&   79.97 / 81.14&73.70 / 73.27&76.83 / 77.20 \\
     Ensemble 1+2+3&   81.69 / 83.62&75.06 / 74.22&\textbf{78.38 / 78.92} \\
    \bottomrule
  \end{tabular}}
\end{table}

\section{Experiments and Results}
\label{sec:exp}

\subsection{Experimental settings}
\label{ssec:exp-setting}

We use the DCASE 2026 Task 7 DIL dataset: three domains presented in sequence over the ten target classes, with $D_1$ audio withheld and $D_2$/$D_3$ audio provided. The per-domain dev-test sets cover class subsets of the ten targets---$D_2$ dev-test has 639 clips (missing baby\_cry, telephone\_ringing) and $D_3$ dev-test has 806 clips (missing knock)---while the released eval set is 3{,}755 hash-named clips with no labels and no domain tags. The training set is heavily imbalanced (speech $\approx 1125$ down to fire $\approx 170$ and baby\_cry $\approx 56$ clips), making the macro metric highly dependent on performance in the minority classes.  All systems share a fixed front end: $32$\,kHz mono audio cropped to $4$\,s, converted to a $64$-bin log-mel spectrogram ($1024$-pt window, $320$-pt hop, $f\!\in\![50,14000]$\,Hz). Because the eval set carries no domain tags, every system runs a single domain-agnostic forward pass per clip---no domain conditioning, and no transductive or test-time augmentation.

\subsection{Implementation details}
\label{ssec:results-systems}

Since the training set is imbalanced, we fix the class-imbalance using \emph{balanced sampling}\cite{psla} for all submitted systems, reshaping which clips populate each batch. The prototype head is trained on the cached penultimate features with Adam (lr $10^{-3}$, batch $64$, $200$ epochs, cosine schedule); its prototypes are centroid-initialized from the cached class means and its temperature $\tau$ is learned. The regression imputer is implemented as a 2-layer MLP with a hidden size of 4096 and trained on D3 rows only (§2.5) using MSE loss for 50 epochs.  

\subsection{Submitted systems and discussion}
\label{ssec:discussion}

Table~\ref{tab:systems} reports the four systems. All share the same prototype classifier recipe (compliant cache $\to$ regression imputation $\to$ balanced sampling $\to$ prototype) and differ only in the backbone. We submit three single-stack systems plus their cross-stack ensemble for robustness against the dev-vs-eval distribution gap.
For context, the official baseline~\cite{mulimani2025dil} scores $\approx 45.5\%$ micro / 53.2\% macro.

All four systems land in a tight $76.8$--$78.4$ micro / $77.2$--$78.9$ macro band on the dev-test average, roughly $+33$~micro and $+25$~macro points above the official checkpoint. The gain is not a single-domain artifact: every system improves both $D_2$ and $D_3$ over the baseline, consistent with the frozen multi-expert design preserving each domain by construction rather than trading one off against the other.

The cross-stack ensemble attains the best dev-test average on \emph{both} micro ($78.38$) and macro ($78.92$). The improvement comes from \emph{backbone diversity}---three genuinely different expert stacks (pure-replay, DeepInversion-replay, and hybrid) making decorrelated errors---rather than from averaging more seeds of one stack.

Across all systems $D_2$ accuracy exceeds $D_3$ by $5$--$8$~micro points. The DeepInversion backbone (System~2) recovers the most $D_3$ accuracy ($75.19$~micro/$73.97$~macro, the best single-system $D_3$), suggesting its data-free generative replay supplies $D_3$-relevant diversity the pure-replay stack lacks; this is why it is retained in the ensemble despite a slightly lower $D_2$ score.

\subsection{Effect of expert diversity on prototype classification}
\label{ssec:results-d5}

Adding only the $E_5$ expert trained with label smoothing and semi-hard negative mining while keeping all other components fixed yields a consistent improvement of approximately 3 percentage points in accuracy across both the pure and inversion-based stacks. In contrast, removing the extra domain experts $E_3$, $E_5$ and retaining only the incremental three-expert stack reduces accuracy to 63.5\%.

These results suggest that the diversity and quality of the expert pool are the primary factors determining prototype classifier performance. The scratch-trained $D_3$ expert provides complementary domain-specific representations that are not fully captured by the incrementally trained experts. While the inversion and feature-imputation mechanisms contribute additional gains, their impact remains comparatively smaller than that of incorporating a strong domain-specialized expert.

\section{Conclusions}
\label{sec:concl}

We treated DIL as a frozen-feature replay problem: stage-wise frozen experts give zero forgetting, and a tiny per-class prototype classifier absorbs all DIL-specific methods. The main factors contributing to the final performance are the diversity of each backbone's expert model and prototype classifier that enables them to collaborate. Cross-stage regression imputation addresses the cross-stage feature-missing problem by estimating expert slots that were unobservable at earlier stages. These results suggest that domain-specific experts, feature imputation, and prototype-based classification provide an effective framework for domain-incremental audio classification under sequential data-access constraints.


\bibliographystyle{IEEEtran}
\bibliography{refs}

\end{sloppy}
\end{document}